%% file: ISWCS18.tex
\documentclass[conference,a4paper]{IEEEtran}

\usepackage[colorlinks = true]{hyperref}

\input{packages_commands_theorems.tex}

\IEEEoverridecommandlockouts
\begin{document}
\title{Information-Theoretic Analysis of D2D-Aided Pipelined Content Delivery in Fog-RAN} 

 \author{%
   \IEEEauthorblockN{Roy Karasik\IEEEauthorrefmark{1},
                     Osvaldo Simeone\IEEEauthorrefmark{2},
                     and Shlomo Shamai (Shitz)\IEEEauthorrefmark{1}}
   \IEEEauthorblockA{\IEEEauthorrefmark{1}%
                     EEE Dept.,
                     Technion - IIT,
                     Technion City,
                     Haifa,
                     Israel}
   \IEEEauthorblockA{\IEEEauthorrefmark{2}%
                     Centre for Telecommunications Research,
                     Dept. of Informatics,
                     King’s College London,
                     UK\\
                 	 \{royk@campus.technion.ac.il, osvaldo.simeone@kcl.ac.uk, sshlomo@ee.technion.ac.il\}}
	\thanks{This work has been supported by the European Research Council (ERC) under the European Union’s Horizon 2020 Research and Innovation Programme (Grant Agreement Nos. 694630 and 725731).} 
 }
\maketitle

\begin{abstract}
	In a Fog-Radio Access Network (F-RAN), edge caching and fronthaul connectivity to a cloud processor are utilized for the purpose of content delivery. Additional Device-to-Device (D2D) communication capabilities can support the operation of an F-RAN by alleviating fronthaul and cloud processing load, and reducing the delivery time. In this work, basic limits on the normalized delivery time (NDT) metric, which captures the high signal-to-noise ratio worst-case latency for delivering any requested content to the users, are derived. Assuming proactive offline caching, out-of-band D2D communication, and an F-RAN with two edge nodes and two users, an information-theoretically optimal caching and delivery strategy is presented. Unlike prior work, the NDT performance is studied under pipelined transmission, whereby the edge nodes transmit on the wireless channel while simultaneously receiving messages over the fronthaul links, and the users transmit messages over the D2D links while at the same time receiving on the wireless channel. Insights are provided on the regimes in which D2D communication is beneficial, and the maximum improvement to the latency is characterized.	
\end{abstract}

\section{Introduction}
For the purpose of content delivery in a cellular system, caching popular data close to clients, i.e., at the Edge Nodes (ENs), is an effective way to reduce delivery time. However, often the cache capacity is not large enough to hold the entire library of popular files, and some part of the requested files has to be downloaded to the ENs on fronthaul links from a remote content server at the cost of additional latency. This added overhead can be reduced by pipelining transmissions on fronthaul and edge channels, i.e., by sending the cached data to the users while simultaneously receiving the rest of requested files from the server \cite{Sen1999proxy}.

Pipelined delivery can also be applied when Device-to-Device (D2D) communication is used to support the operation of content delivery. This is especially useful when out-of-band D2D links, whereby direct communication between the users takes place over frequency resources that are orthogonal with respect to the spectrum used for cellular transmission, are available \cite{le2016microcast}.

\begin{figure}[!t]
	\centering
	\includegraphics[width=\columnwidth]{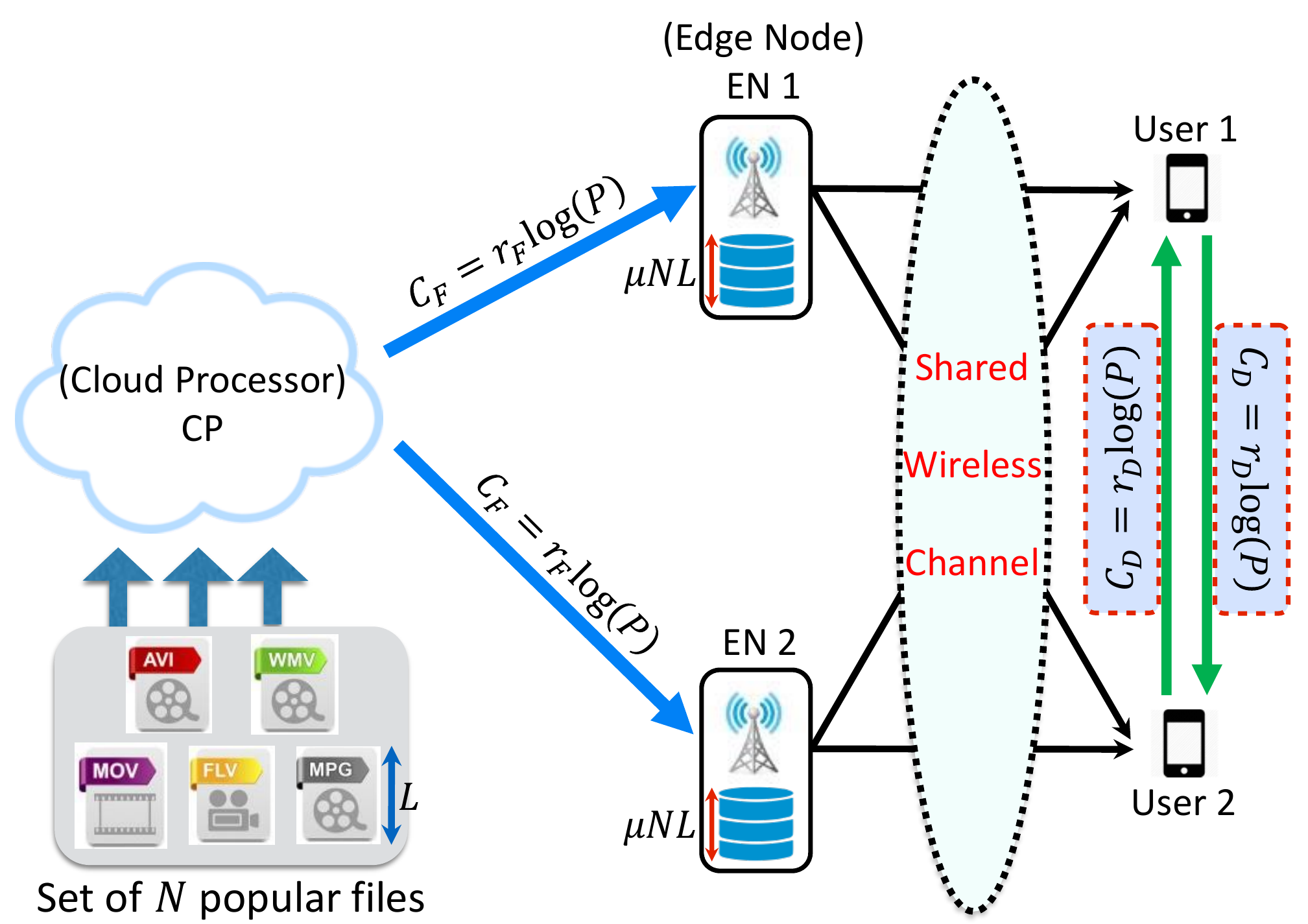}
	\caption{Illustration of the D2D-aided F-RAN model under study.}
	\label{fig_model}
\end{figure}

In this work, we study the impact of pipelining fronthaul and D2D communications on content delivery time for a Fog-Radio Access Network (F-RAN), in which, as illustrated in Fig \ref{fig_model}, both edge caching and fronthaul connectivity to a Cloud Processor (CP) are leveraged \cite{hung2015architecture,tandon2016harnessing}. 

\textbf{Related Work:} The Normalized Delivery Time (NDT) is a metric which measures the high signal-to-noise ratio (SNR) worst-case latency relative to an ideal system with unlimited caching capability, and was introduced in \cite{sengupta2016cache} for cache-aided interference channels. The NDT of a general F-RAN system with pipelined fronthaul-edge transmission was studied in \cite{sengupta2016pipelined, sengupta2016fog}, where the proposed schemes were shown to achieve the minimum NDT to within a factor of 2, and the minimum NDT was completely characterized for two ENs and two users, as well as for other special cases. The $2\times2$ D2D-aided F-RAN system with serial delivery policies was analyzed in \cite{karasik2018ISIT}, and an optimal strategy for utilizing the fronthaul and D2D links, as well as the downlink wireless channel, was shown to achieve the minimum NDT. The strategy was based on a novel scheme for the X-channel with receiver cooperation. A cache-aided interference channel where both ENs and users have caching capabilities was studied in \cite{naderializadeh2017fundamental}, and lower and upper bound on the sum degrees-of-freedom were presented under the constraint of linear precoders at the ENs. 
Content delivery in a D2D caching network, i.e., without ENs, was studied in \cite{jeon2017wireless}, where the per-node capacity scaling law was characterized for multi-hop D2D communication.

\textbf{Main Contributions:} In this paper, we characterize the minimum NDT for the D2D-aided F-RAN illustrated in Fig. \ref{fig_model} under pipelined delivery. The minimum NDT is used to identify the conditions under which D2D communication is beneficial, and to provide insights on the interplay between fronthaul and D2D resources. Furthermore, using the results for serial transmission presented in \cite{karasik2018ISIT}, we identify the improvement to the NDT that can be achieved by pipelining fronthaul and D2D transmissions. 

\section{System Model}\label{sec:sys_model}
We consider content delivery via the Fog-Radio Access Network (F-RAN) system with Device-to-Device (D2D) links depicted in Fig. \ref{fig_model}. 
Each user requests a file from a given library, and the goal is to deliver the requested files to the users within the lowest possible delivery latency. To this end, the two single-antenna users receive symbols from the two single-antenna Edge Nodes (ENs) over a downlink wireless channel, and can cooperate by utilizing two orthogonal out-of-band D2D links. Each EN is connected to a Cloud Processor (CP) by a fronthaul link.

Formally, Let $\mathcal F$ denote a library of $N\geq 2$ files, $\mathcal F=\{f_1,\ldots,f_N\}$, each of size $L$ bits. The entire library is available at the CP, whereas the ENs can only store up to $\mu NL$ bits each, where $0\leq\mu\leq 1$ is the fractional cache size. The library is fixed for the considered time period. The operation of the system is divided into two phases: placement and delivery. During the placement phase, contents are proactively cached at the ENs, subject to the mentioned cache capacity constraints.

After the placement phase, the system enters the delivery phase, which is organized in Transmission Intervals (TIs). In every TI, each user arbitrarily requests one of the $N$ files from the library. The users' requests in a given TI are denoted by the demand vector $\mathbf d\triangleq (d_1,d_2)\in[N]^2$, where for any positive integer $a$, we define the set $[a]\triangleq\{1,2,\ldots,a\}$. This vector is known at the beginning of a TI at the CP and ENs.

Denote the duration of each TI by $T$ symbols. At time $t\in[T]$, each user $k\in[2]$ receives a channel output given by 
\begin{IEEEeqnarray}{rCl}\label{eq:wireless_channel}
	y_k[t]&=&h_{k1}x_1[t]+h_{k2}x_2[t]+z_k[t],
\end{IEEEeqnarray}
where $x_m[t]\in\mathbb C$ is the baseband symbol transmitted from EN $m\in[2]$ at time $t$, which is subject to the average power constraint $\mathbb E |x_m(t)|^2\leq P$ for some $P>0$; coefficient $h_{km}\in\mathbb C$ denotes the quasi-static flat-fading channel between EN $m$ to user $k$, which is assumed to be drawn i.i.d. from a continuous distribution and remain constant during each TI; and $z_k[t]$ is an additive white Gaussian noise, such that $z_k[t]\sim\mathcal C\mathcal N(0,1)$ is independent and identically distributed (i.i.d.) across time and users. The Channel State Information (CSI) $\mathbf{H}\triangleq\{h_{km}:k\in[2],m\in[2]\}$ is known to all nodes.

Furthermore, at time $t\in[T]$, EN $m$ receives a message $u_m[t]$ from the CP over a fronthaul link of capacity $C_F=r_F\log (P)$ bits per symbol. With this parametrization, the fronthaul rate $r_F\geq 0$ represents the ratio between the fronthaul capacity and the high-SNR capacity of each EN-to-user wireless link in the absence of interference. In addition, the users can cooperate by sending messages $v_1[t]$ and $v_2[t]$ from user 1 and user 2, respectively, over orthogonal D2D links of capacity $C_D=r_D\log (P)$ bits per symbol.

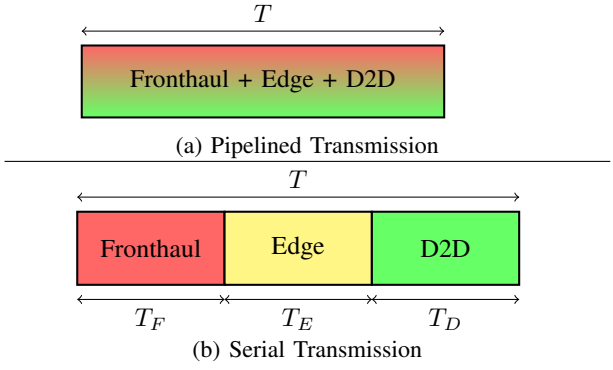
\begin{figure}[!t]
	\centering
	\begin{subfigure}[!t]{\columnwidth}
		\centering
		\resizebox {0.7\columnwidth} {!} {	
			\input{tikz_pipelined_tx.tex}	
		}
		\caption{Pipelined Transmission}
		\label{fig:TI_pipelined}
	\end{subfigure}
	\rule{0.9\columnwidth}{.4pt}
	\begin{subfigure}[!t]{\columnwidth}
		\centering
		\resizebox {0.7\columnwidth} {!} {	
			\input{tikz_serial_tx.tex}
		}
		\vspace{-0.5\baselineskip}	
		\caption{Serial Transmission}
		\label{fig:TI_serial}
	\end{subfigure}
	\caption{Transmission Interval structure for either serial or pipelined delivery policies.}
\end{figure}

Notice that, in the system described above, as illustrated in Fig. \ref{fig:TI_pipelined}, the ENs can simultaneously receive messages over the fronthaul links and transmit on the wireless channel; and the users can receive on the wireless channel while, at the same time, transmitting messages on the D2D links. Following \cite{sengupta2016fog}, we refer to this model as enabling pipelined delivery.
An alternative model with serial delivery was considered in \cite{karasik2018ISIT}, where, as seen in Fig. \ref{fig:TI_serial}, in each TI, the CP first sends the fronthaul messages to the ENs for a total time of $T_F$ symbols; then, the ENs transmit on the wireless shared channel for a total time of $T_E$ symbols; and, finally, the users use the out-of-band D2D links for a total time of $T_D$ symbols.

\subsection{Caching, Delivery and D2D Transmission}
The operation of the system is defined by policies that perform caching, as well as delivery via fronthaul, edge and D2D communication resources.
\subsubsection{Caching Policy} During the placement phase, for EN $m$, $m\in[2]$, the caching policy is defined by functions $\pi^m_{c,n}(\cdot)$ that map each file $f_n$ to its cached content $s_{m,n}$ as
\begin{IEEEeqnarray}{rCl'l}\label{eq:caching_policy}
	s_{m,n}&=& \pi_{c,n}^m(f_n),&\forall n\in[N].
\end{IEEEeqnarray}
Note that, as per \eqref{eq:caching_policy}, we consider policies where only coding within each file is allowed, i.e., no inter-file coding is permitted. The cached content $s_{m,n}$ has to satisfy the cache capacity constraint, and hence we have $H(s_{m,n})\leq \mu L$. The overall cache content at EN $m$ is given by $s_m\triangleq(s_{m,1},s_{m,2}\ldots,s_{m,N})$.
\subsubsection{Fronthaul Policy}In each TI of the delivery phase, for EN $m$, $m\in[2]$, the CP maps the library $\mathcal F$, the demand vector $\mathbf{d}$ and CSI $\mathbf{H}$ to the fronthaul message
\begin{IEEEeqnarray}{rCl}
	\mathbf{u}_m=(u_m[1],u_m[2],\ldots,u_m[T])=\pi_f^m(\mathcal F,s_m,\mathbf{d},\mathbf{H}).
\end{IEEEeqnarray}
Note that the fronthaul message cannot exceed $TC_F$ bits, i.e., $H(\mathbf{u}_m)\leq Tr_F\log (P)$.
\subsubsection{Edge Transmission Policies}Each EN $m$ at time $t\in[T]$ uses a function $\pi_{e,t}^m(\cdot)$ to map the local cache content, the fronthaul messages received up to time $t-1$, the demand vector, and the global CSI to the output symbol
\begin{IEEEeqnarray}{c}
	x_m[t]=\pi_{\text{e},t}^m(s_m,u_m[1],u_m[2],\ldots,u_m[t-1],\mathbf{d},\mathbf{H}).
\end{IEEEeqnarray}
\subsubsection{D2D Interactive Communication Policies}At time instant $t\in[T]$, user $k$ transmits using the function $\pi^k_{\text{D2D},t}(\cdot)$ that maps the received edge signal up to time $t-1$, global CSI, and the previously received D2D message from user $k'\neq k$ to the D2D message
\begin{IEEEeqnarray}{rl}
	v_k[t]=\pi_{\text{D2D},t}^k&\left(y_k[1],\ldots,y_k[t-1],\mathbf{H},v_{k'}[1],\ldots,v_{k'}[t-1],\right.\IEEEnonumber\\
	&\quad v_{k}[1],\ldots,v_{k}[t-1]).
\end{IEEEeqnarray}
The total size of each D2D message cannot exceed $TC_D$ bits, i.e., $H(\mathbf{v}_{k})\leq Tr_D\log(P)$, where $\mathbf{v}_{k}\triangleq(v_{k}[1],\ldots,v_{k}[T])$.
\subsubsection{Decoding Policy}At the end of each TI, user $k$ implements a decoding policy $\pi_d^k(\cdot)$ that maps the channel outputs, the D2D messages from user $k'\neq k\in[2]$, the user demand and the global CSI to an estimate of the requested file $f_{d_k}$ given as
\begin{IEEEeqnarray}{rCl}
	\hat{f}_{d_k}&=&\pi_d^k(\mathbf{y}_k,\mathbf{v}_{k'},d_k,\mathbf{H}),
\end{IEEEeqnarray}
where $\mathbf{y}_k\triangleq(y_k[1],\ldots,y_k[T])$.

The probability of error is defined as
\begin{IEEEeqnarray}{rCl}
	P_e&\triangleq&\max_{\mathbf{d}}\max_{k\in[2]}\Pr(\hat{f}_{d_k}\neq f_{d_k}),
\end{IEEEeqnarray}
which is the worst-case probability of decoding error measured over all possible demand vectors $\mathbf{d}$ and over all users. A sequence of policies, indexed by the file size $L$, is said to be feasible if, for almost all channel realization $\mathbf{H}$, we have $P_e\rightarrow0$ when $L\rightarrow\infty$.

\subsection{Performance Metric}
We adopt the Normalized Delivery Time (NDT), introduced in \cite{sengupta2016fog}, as the performance metric of interest. The NDT is a relative measure used to evaluate the performance in the high-SNR regime. It is defined as the ratio between the worst-case delivery time per bit required to satisfy any possible demand vector $\mathbf{d}$ and the delivery time per bit for an ideal reference system with unlimited cache capacity and no interference. 

Accordingly, for any sequence of feasible policies, we define the NDT under pipelined delivery as
\begin{IEEEeqnarray}{rCl}\label{eq:NDT_def_P}
	\delta_\text{P}(\mu,r_F,r_D)&\triangleq&\lim_{P\rightarrow\infty}\lim_{L\rightarrow\infty}\frac{\mathbb E[T]}{L/\log(P)},
\end{IEEEeqnarray}
where the notation emphasizes the dependence of the NDT on the fractional cache size $\mu$, and the fronthaul and D2D rates $r_F$ and $r_D$, respectively. The factor $L/\log(P)$, used for normalizing the delivery time in \eqref{eq:NDT_def_P}, represents the minimal time to deliver a file in the reference system mentioned above.

The minimum NDT under pipelined delivery is finally defined as the minimum over all achievable policies
\begin{IEEEeqnarray}{l}\label{eq:minimun_pipelined_ndt_def}
	\delta^*_\text{P}(\mu,r_F,r_D)\triangleq\IEEEnonumber\\
	\qquad\inf\{\delta_\text{P}(\mu,r_F,r_D):\delta_\text{P}(\mu,r_F,r_D)\text{ is achievable}\}.
\end{IEEEeqnarray} 
By construction, we have the lower bound $\delta^*(\mu,r_F,r_D)\geq 1$.

Similarly, for the serial delivery policies studied in \cite{karasik2018ISIT}, the fronthaul, edge and D2D NDTs are defined as
\begin{IEEEeqnarray}{rcl}
	\delta_F&\triangleq&\lim_{P\rightarrow\infty}\lim_{L\rightarrow\infty}\frac{\mathbb E[T_F]}{L/\log(P)},\label{eq:F_NDT}\\
	\delta_E&\triangleq&\lim_{P\rightarrow\infty}\lim_{L\rightarrow\infty}\frac{\mathbb E[T_E]}{L/\log(P)}\label{eq:E_NDT}
\end{IEEEeqnarray}
and
\begin{IEEEeqnarray}{c}
	\delta_D\triangleq\lim_{P\rightarrow\infty}\lim_{L\rightarrow\infty}\frac{\mathbb E[T_D]}{L/\log(P)},\label{eq:D_NDT}
\end{IEEEeqnarray}
respectively, and the minimum NDT is defined as
\begin{IEEEeqnarray}{l}\label{eq:minimum_ndt_def}
	\delta^*(\mu,r_F,r_D)\triangleq\IEEEnonumber\\
	\qquad\inf\{\delta(\mu,r_F,r_D):\delta(\mu,r_F,r_D)\text{ is achievable}\},
\end{IEEEeqnarray} 
where $\delta(\mu,r_F,r_D)\triangleq\delta_F+\delta_E+\delta_D$.

\section{NDT Analysis for Pipelined Delivery}\label{sec:pipelined}
In this section, we study the D2D-aided F-RAN model with pipelined delivery as defined in Sec. \ref{sec:sys_model}. The minimum pipelined NDT is characterized, and conclusions are drawn on the role of D2D communication in improving the delivery latency and in reducing the requirements on fronthaul and cache resources. We also compare the performance of optimal pipelined and serial delivery policies.

\subsection{Minimum NDT}\label{sec:min_NDT_pipelined}
The following Lemma extends \cite[Lemma 4]{sengupta2016fog} in order to bound the improvement in NDT that can be achieved by the use of pipelined delivery policies.
\begin{lemma}\label{lem:loose_bound}
	For the $2\times 2$ F-RAN system, pipelined transmission can reduce the minimum NDT, as compared to serial transmission, by a factor of at most three, i.e.,
	\begin{IEEEeqnarray}{c}\label{eq:loose_bound}
		\frac{\delta^*(\mu,r_F,r_D)}{\delta^*_\text{P}(\mu,r_F,r_D)}\leq3.
	\end{IEEEeqnarray}
\end{lemma}
\begin{IEEEproof}
	Akin to the proof of \cite[Lemma 4]{sengupta2016fog}, consider an optimal pipelined delivery policy that achieves the minimum NDT $\delta^*_\text{P}(\mu,r_F,r_D)$. A serial delivery policy can be constructed from it by using the same strategy for fronthaul, edge and D2D transmissions, with the caveat that the ENs start transmitting only after the fronthaul transmission is complete; and the users start conferencing after receiving all the symbols on the shared wireless channel. The resulting serial NDT is hence three times that of the minimum pipelined NDT, and is also an upper bound on the minimum NDT under serial delivery, i.e., $\delta^*(\mu,r_F,r_D)\leq 3\delta^*_\text{P}(\mu,r_F,r_D).$ 
\end{IEEEproof}

While Lemma \ref{lem:loose_bound} provides only an upper bound, the following theorem fully characterizes the minimum NDT in the presence of pipelined delivery.
\begin{theorem}\label{th:minimum_NDT_pipelined}
	The minimum NDT for the $2\times 2$ F-RAN system under pipelined delivery with number of files $N\geq 2$, fractional cache size $\mu\geq0$, fronthaul rate $r_F\geq 0$ and D2D rate $r_D\geq 0$ is given as
	\begin{IEEEeqnarray}{c}\label{eq:minimum_NDT_pipelined}
		\delta^*_\text{P}(\mu,r_F,r_D)=\max\cb{\frac{1-2\mu}{r_F},\frac{2-\mu}{1+r_F+r_D},1}.
	\end{IEEEeqnarray}
\end{theorem}
\begin{IEEEproof}
	The proof is omitted for brevity and is provided in \cite[Appendix A]{karasik2018information}.
\end{IEEEproof}

\begin{figure}[!t]
	\centering
	\resizebox{\columnwidth} {!} {	
	\input{tikz_blockMarkov.tex}
	}
	\caption{Pipelining via block-Markov encoding.}
	\label{fig:block-Markov}
\end{figure}
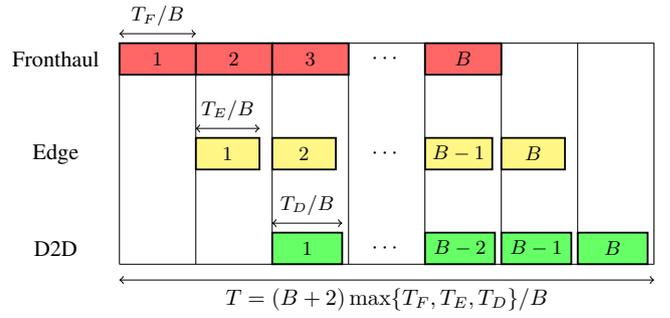

The policies that achieve the minimum NDT \eqref{eq:minimum_NDT_pipelined} are based on a block-Markov encoding technique that converts a serial delivery policy into a pipelined delivery policy. The approach is a generalization of the method presented in \cite[Sec. VI.B]{sengupta2016fog} for an F-RAN with no D2D links. To elaborate, fix a serial delivery policy with its fronthaul, edge and D2D transmission strategy. As illustrated in Fig. \ref{fig:block-Markov}, in order to convert this strategy into one that leverages pipelining, every file in the library is split into $B$ blocks of size $L/B$ bits each, and every TI is divided into $B+2$ slots. In each slot $b\in[B]$, the CP uses the fronthaul links to deliver the $b$th block of the requested files using the fronthaul transmission strategy of the selected serial policy. At the same time, the ENs, having received the fronthaul message for the $(b-1)$th block in the previous slot, apply the edge transmission strategy of the serial policy to deliver the $(b-1)$th block of the requested files to the users; and the users apply the corresponding conferencing scheme to cooperate in the decoding of the $(b-2)$th block of the requested files. For a serial delivery scheme that achieves fronthaul, edge and D2D transmission durations $T_F$, $T_E$ and $T_D$, respectively, the block-Markov approach, with arbitrarily large number of blocks $B$, achieves the pipelined NDT 
\begin{IEEEeqnarray}{rCl}\label{eq:block-Markov}
	\delta_\text{P,ach}(\mu,r_F,r_D)&=&\lim_{B\rightarrow\infty}\lim_{P\rightarrow\infty}\lim_{L\rightarrow\infty}\frac{B+2}{B}\cdot\frac{\max\{T_F,T_E,T_D\}}{L/\log(P)}\IEEEnonumber\\
	&=&\max\{\delta_F,\delta_E,\delta_D\},
\end{IEEEeqnarray}
where $\delta_F$, $\delta_E$ and $\delta_D$ are the fronthaul, edge and D2D NDTs of the serial transmission scheme as defined in \eqref{eq:F_NDT}, \eqref{eq:E_NDT} and \eqref{eq:D_NDT}, respectively.

The result in Theorem \ref{th:minimum_NDT_pipelined} can be leveraged to draw conclusions on the role of D2D cooperation in improving the delivery latency. 
In this regard, we note that, for $r_F\geq 1$, the minimum NDT \eqref{eq:minimum_NDT_pipelined} is given by $\delta^*_\text{P}(\mu,r_F,r_D)=1$ for all $\mu\geq0$ and $r_D\geq 0$. It follows that, with a sufficiently large fronthaul capacity, the ideal NDT can be achieved without requiring any caching and D2D resources at the edge (see also \cite[Sec. VI.D]{sengupta2016fog}). Furthermore, for any $r_F<1$ and sufficiently small or large fractional cache size $\mu$, i.e., for $\mu< (1-r_F)/(2+r_F)$ or $\mu\geq (1-r_F)$, the minimum NDT is identical to the minimum NDT without D2D links ($r_D=0$) derived in \cite[Corollary 5]{sengupta2016fog}. To sum up, D2D communication provides a latency reduction only when $r_F<1$ and the cache capacity takes values in the intermediate regime $(1-r_F)/(2+r_F)\leq\mu<(1-r_F)$. 

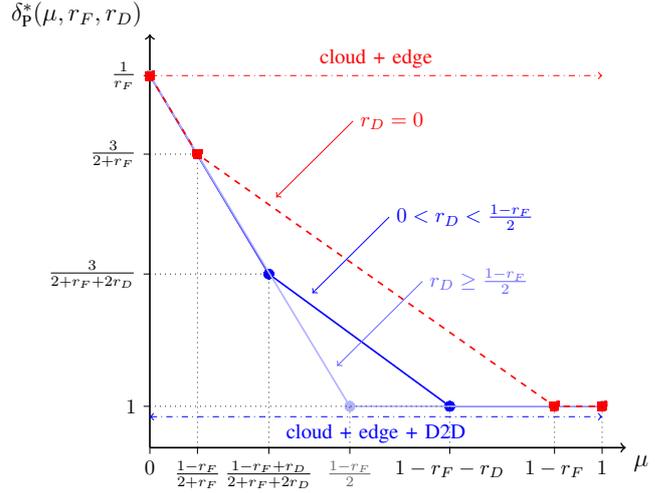
\begin{figure}[!t]
	\centering
	\resizebox {\columnwidth} {!} {
		\input{tikzNDT_pipe1.tex}
	}
	\caption{Minimum pipelined NDT for the $2\times 2$ D2D-aided F-RAN as a function of $\mu$ with $r_F<1$. Note that the positions of $\mu=1-r_F-r_D$ and $\mu=(1-r_F)/2$ are indicated for $r_D<(1-r_F)/2$ and $r_D\geq(1-r_F)/2$, respectively.}
	\label{fig:NDT_pipelined}
\end{figure}

The discussion above is illustrated in Fig. \ref{fig:NDT_pipelined}, where we plot the minimum NDT \eqref{eq:minimum_NDT_pipelined} as a function of the fractional cache size $\mu$ for fixed fronthaul rate $r_F<1$ and D2D rate $r_D$. 
For small cache capacities satisfying the inequality $\mu< (1-r_F)/(2+r_F)$, D2D communication cannot reduce the minimum NDT because, in this regime, the latency bottleneck (cf. \eqref{eq:block-Markov}) is caused by fronthaul communication, which is required to deliver a large part of the requested files. Conversely, for $\mu\geq 1-r_F$, the cache capacity is large enough to support delivery via cache-aided cooperative Zero-Forcing (ZF) \cite[Lemma 2]{sengupta2016fog} with a fronthaul overhead that does not affect the achievability of the ideal NDT of one. Fig. \ref{fig:NDT_pipelined} also illustrates that, thanks to pipelining, the ideal NDT of one can be obtained even when the cache capacity is not sufficient to store the entire library. This conclusion contrasts with serial delivery \cite{karasik2018ISIT}.

\begin{figure}[!t]
	\centering
	\includegraphics[width=0.9\columnwidth]{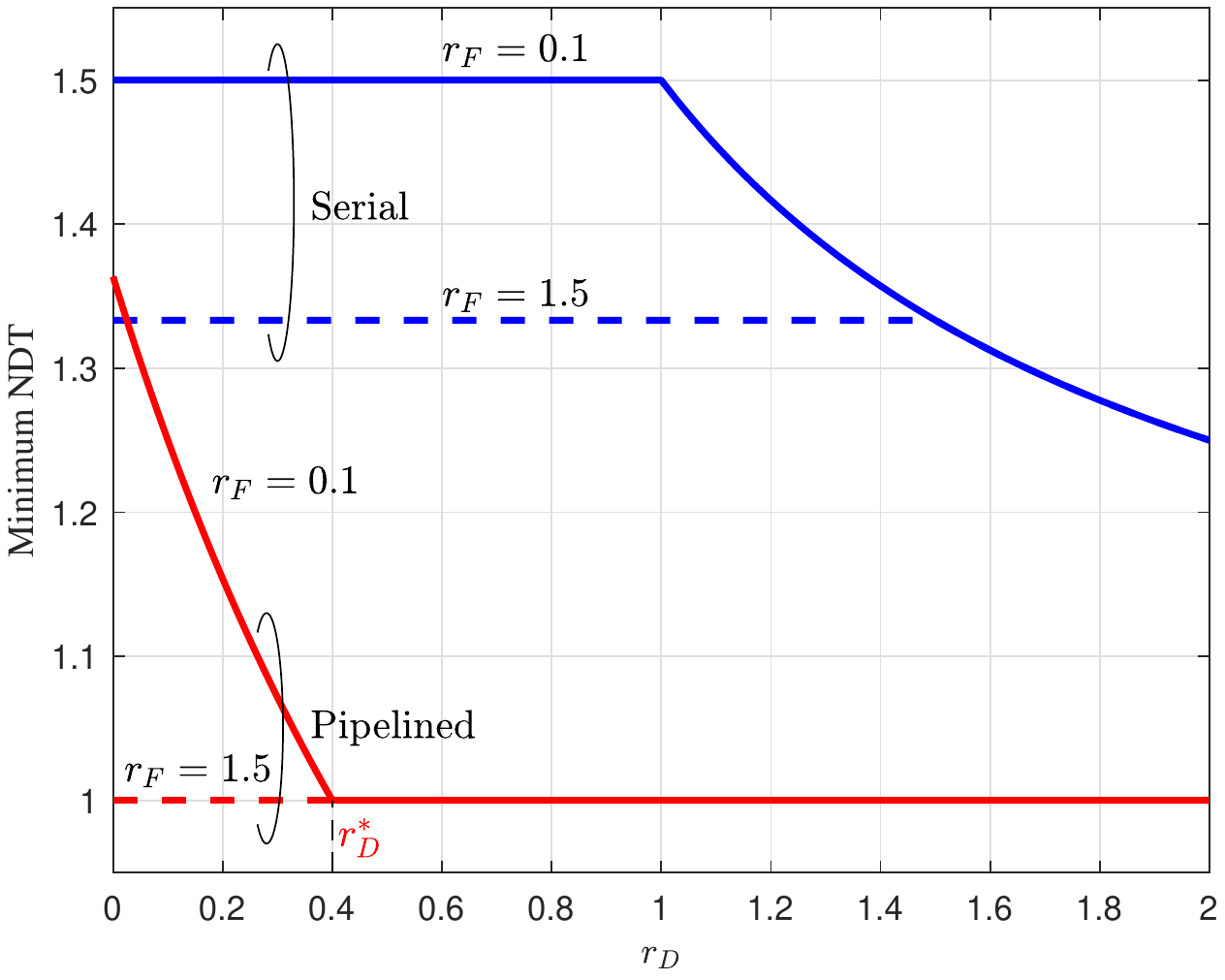}
	\caption{Minimum NDT for the $2\times 2$ D2D-aided F-RAN as a function of $r_D$ for $\mu=0.5$.}
	\label{fig:NDT_vs_rD}
\end{figure}

As another important observation, unlike serial delivery policies, where the minimum NDT is a strictly decreasing function of $r_D$ for all $r_D>\max\{1,r_F\}$ and $0<\mu<1$ \cite[Theorem 1]{karasik2018ISIT}, with pipelined delivery, the minimum NDT \eqref{eq:minimum_NDT_pipelined} is a constant function of $r_D$ for $r_D\geq r_D^*(\mu,r_F)$, where
\begin{IEEEeqnarray}{rl}\label{eq:rD_threshold}
	r^*_D&(\mu,r_F)\triangleq\\
	&\left\lbrace\begin{array}{ll}
		\min\cb{1-r_F-\mu,\frac{r_F(1+\mu)}{1-2\mu}-1}&\text{for }\mu<1/2\\
		1-r_F-\mu&\text{for }\mu\geq 1/2.
	\end{array}\right.\IEEEnonumber
\end{IEEEeqnarray} 
This is because, when $r_D\geq r_D^*(\mu,r_F)$, the duration of the D2D transmission in each slot of the optimal block-Markov strategy is smaller than the fronthaul or edge transmissions (cf. Fig. \ref{fig:block-Markov}). Hence, as per \eqref{eq:block-Markov}, increasing the D2D rate $r_D$ beyond $r_D^*(\mu,r_F)$ does not reduce the minimum NDT. 
This is illustrated in Fig. \ref{fig:NDT_vs_rD}, where we plot the minimum NDT, under either serial or pipelined delivery, as a function of the D2D rate $r_D$, for a fixed fractional cache size $\mu=1/2$ (The graphs for serial delivery are based on \cite[Theorem 1]{karasik2018ISIT}).
Fig. \ref{fig:NDT_vs_rD} illustrates that, for serial transmission, an arbitrarily large D2D rate $r_D$ is required for the D2D overhead to become negligible and to achieve a minimum NDT of one. In contrast, with pipelined transmission, increasing the D2D rate $r_D$ beyond the threshold value $r^*_D(\mu,r_F)$ does not affect the minimum NDT. 
In addition, for $r_F\geq 1$, we have $r^*_D(\mu,r_F)\leq 0$, and the minimum pipelined NDT can be achieved without utilizing the D2D links.
The figure also highlights the gains that can be achieved by means of pipelining.

\subsection{Latency Reduction via Pipelining}\label{sec:pipe_gain}
The following corollary characterizes the latency reduction that can be achieved by pipelining fronthaul and D2D transmissions.
\begin{corollary}\label{cor:pipe_gain}
	For the $2\times 2$ F-RAN system, the pipelining gain $\delta^*(\mu,r_F,r_D)/\delta_\text{P}^*(\mu,r_F,r_D)$ satisfies the following:
	\begin{itemize}
		\item \textit{Low cache capacity} ($0\leq\mu<1/2$): \\$\delta^*(\mu,r_F,r_D)/\delta_\text{P}^*(\mu,r_F,r_D)\leq2+\mu$, and the bound is achieved with equality under the conditions $r_F=1-2\mu$ and $\mu\leq r_D\leq 1$;
		\item \textit{High cache capacity} ($1/2\leq\mu\leq 1$): \\$\delta^*(\mu,r_F,r_D)/\delta_\text{P}^*(\mu,r_F,r_D)\leq2-\mu$, and the bound is achieved with equality under the conditions $r_F,r_D\leq 1$ and $r_F+r_D\geq 1-\mu$.
	\end{itemize}
\end{corollary}
\begin{IEEEproof}
	The corollary follows directly by dividing $\delta^*(\mu,r_F,r_D)$ in \cite[Theorem 1]{karasik2018ISIT} by $\delta_\text{P}^*(\mu,r_F,r_D)$ in \eqref{eq:minimum_NDT_pipelined}.
\end{IEEEproof}

Corollary \ref{cor:pipe_gain} implies that, by pipelining both fronthaul and D2D transmissions, the NDT can be reduced by a factor of $2.5$. While being smaller than the simple upper bound in Lemma \ref{lem:loose_bound}, this result highlights the potential advantages of pipelining. In addition, it follows from Corollary \ref{cor:pipe_gain} that, when $r_F=0$, i.e., when no fronthaul communication is enabled, the maximum gain of pipelining D2D communication is $1.5$, and is achieved for $\mu=1/2$ and $1/2\leq r_D\leq 1$. Conversely, when $r_D=0$, i.e., when there are no D2D links, the maximum gain of pipelining fronthaul transmission is $2$, and is achieved for $\mu=0$ and $r_F=1$.

As final observation on Corollary \ref{cor:pipe_gain}, for $\mu>1/2$, some part of the requested files can be delivered with minimal latency using cache-aided cooperative ZF, i.e., without utilizing the fronthaul and D2D links. As a result, the maximum pipelining gain decreases with $\mu$. In contrast, for $\mu<1/2$, the maximum pipelining gain increases with $\mu$, since, by increasing the cache capacity $\mu$, an ideal NDT of one can be achieved with smaller fronthaul rate $r_F$. This is because a larger part of the requested files can be delivered using the D2D-based scheme presented in \cite[Sec. III]{karasik2018ISIT}.

\section{Conclusions}
In this work, information-theoretic insights were provided on the benefits of D2D communication for content delivery in an F-RAN with pipelined transmission policies. 
The minimum normalized delivery time was characterized, and it was shown that, in contrast to serial transmission, a pipelined delivery strategy with relatively small D2D capacity can achieve an ideal interference-free latency. Furthermore, pipelining both fronthaul and D2D transmissions can improve the delivery time as compared to serial transmission by a factor of $2.5$.  
Among possible extensions of this work, we mention the generalization of the results to more than two users and ENs, and the evaluation of the impact of imperfect CSI.

\bibliographystyle{IEEEtran}
\bibliography{IEEEabrv,myBib}

\end{document}

%% file: packages_commands_theorems.tex
\usepackage{cite}

\ifCLASSINFOpdf
 \usepackage[pdftex]{graphicx}
 \graphicspath{{../pdf/}{../jpeg/}}
 \DeclareGraphicsExtensions{.pdf,.jpeg,.png}
\else
 \usepackage[dvips]{graphicx}
 \graphicspath{{../eps/}}
 \DeclareGraphicsExtensions{.eps}
\fi

\usepackage[cmex10]{amsmath}  
\usepackage{amssymb}
\usepackage{algorithmic}

\usepackage{array}

\usepackage{caption}
\usepackage{subcaption}	

\usepackage{stfloats}

\usepackage{flafter}

\usepackage{url}

\usepackage{cases}

\usepackage{tikz}
\usetikzlibrary{calc} 
\usepackage{pgfplots}
\usetikzlibrary{patterns}

\newcommand{\cb}[1] {\left\lbrace #1 \right\rbrace}

\newtheorem{lemma}{Lemma}

\newtheorem{theorem}{Theorem}
\newtheorem{corollary}{Corollary}

%% file: tikz_pipelined_tx.tex
\begin{tikzpicture}

\node[thick,draw,minimum width=5cm,minimum height=1cm,top color=red!60,bottom color=green!60] at (0,0) {Fronthaul + Edge + D2D};
\node at (3.5,0) {}; 

\path[draw,<->] (-2.5,0.7) -- node[midway,above]{$T$} (2.5,0.7);
	
\end{tikzpicture}

%% file: tikz_serial_tx.tex
\begin{tikzpicture}

\node[thick,draw,minimum width=2cm,minimum height=1cm,fill=red!60] at (0,0) {Fronthaul};
\node[thick,draw,minimum width=2cm,minimum height=1cm,fill=yellow!60] at (2,0) {Edge};
\node[thick,draw,minimum width=2cm,minimum height=1cm,fill=green!60] at (4,0) {D2D};

\path[draw,<->] (-1,0.7) -- node[midway,above]{$T$} (5,0.7);
\path[draw,<->] (-1,-0.7) -- node[midway,below]{$T_F$} (1,-0.7);
\path[draw,<->] (1,-0.7) -- node[midway,below]{$T_E$} (3,-0.7);
\path[draw,<->] (3,-0.7) -- node[midway,below]{$T_D$} (5,-0.7);
	
\end{tikzpicture}

%% file: tikz_blockMarkov.tex
\begin{tikzpicture}
\node[minimum width=1.5cm] at (0,0) {Fronthaul};
\node[minimum width=1.5cm] at (0,-1.5) {Edge};
\node[minimum width=1.5cm] at (0,-3) {D2D};
\node[thick,draw,minimum width=1.2cm,minimum height=0.5cm,fill=red!60] at (1.6,0) {\small$1$};
\node[thick,draw,minimum width=1.2cm,minimum height=0.5cm,fill=red!60] at (2.8,0) {\small$2$};
\node[thick,draw,minimum width=1.2cm,minimum height=0.5cm,fill=red!60] at (4,0) {\small$3$};
\node[minimum width=1.2cm,minimum height=0.5cm] at (5.2,0) {$\cdots$};
\node[minimum width=1.2cm,minimum height=0.5cm] at (5.2,-1.5) {$\cdots$};
\node[minimum width=1.2cm,minimum height=0.5cm] at (5.2,-3) {$\cdots$};
\node[thick,draw,minimum width=1.2cm,minimum height=0.5cm,fill=red!60] at (6.4,0) {\small$B$};
\node[thick,draw,minimum width=1.1cm,minimum height=0.5cm,fill=green!60] at (3.95,-3) {\small$1$};
\node[thick,draw,minimum width=1.1cm,minimum height=0.5cm,fill=green!60] at (6.35,-3) {\small$B-2$};
\node[thick,draw,minimum width=1.1cm,minimum height=0.5cm,fill=green!60] at (7.55,-3) {\small$B-1$};
\node[thick,draw,minimum width=1.1cm,minimum height=0.5cm,fill=green!60] at (8.75,-3) {\small$B$};
\node[thick,draw,minimum width=1cm,minimum height=0.5cm,fill=yellow!60] at (2.7,-1.5) {\small$1$};
\node[thick,draw,minimum width=1cm,minimum height=0.5cm,fill=yellow!60] at (3.9,-1.5) {\small$2$};
\node[thick,draw,minimum width=1cm,minimum height=0.5cm,fill=yellow!60] at (6.33,-1.5) {\small $B-1$};
\node[thick,draw,minimum width=1cm,minimum height=0.5cm,fill=yellow!60] at (7.5,-1.5) {\small$B$};

\path[draw,<->] (1,0.4) -- node[midway,above]{\small $T_F/B$} (2.2,0.4);
\path[draw,<->] (2.2,-1.1) -- node[midway,above]{\small $T_E/B$} (3.2,-1.1);
\path[draw,<->] (3.4,-2.6) -- node[midway,above]{\small $T_D/B$} (4.5,-2.6);
\path[draw,<->] (1,-3.5) -- node[midway,below]{$T=(B+2)\max\{T_F,T_E,T_D\}/B$} (9.4,-3.5);

\path[draw] (1,0.25) -- (1,-3.25);
\path[draw] (2.2,0.25) -- (2.2,-3.25);
\path[draw] (3.4,0.25) -- (3.4,-3.25);
\path[draw] (4.6,0.25) -- (4.6,-3.25);
\path[draw] (5.8,0.25) -- (5.8,-3.25);
\path[draw] (7,0.25) -- (7,-3.25);
\path[draw] (8.2,0.25) -- (8.2,-3.25);
\path[draw] (9.4,0.25) -- (9.4,-3.25);
\path[draw] (1,0.25) -- (9.4,0.25);
\path[draw] (1,-3.25) -- (9.4,-3.25);

\end{tikzpicture}

%% file: tikzNDT_pipe1.tex
\begin{tikzpicture}
	\coordinate (origin) at (0,0);
	\coordinate (top) at (0,7);
	\coordinate (right) at (8,0);
	
	\draw[thick,->] (origin) -- (top) node[anchor=south east]{\large $\delta_\text{P}^*(\mu,r_F,r_D)$};
	\draw[thick,->] (origin) -- (right) node[anchor=north west]{\large $\mu$};
	
	\coordinate (X1) at ($(origin)!0.95!(right)$); 	
	\coordinate (X2) at ($(origin)!0.85!(right)$);	
	\coordinate (X3) at ($(origin)!0.63!(right)$);	
	\coordinate (X4) at ($(origin)!0.25!(right)$);	
	\coordinate (X5) at ($(origin)!0.1!(right)$);	
	\coordinate (X6) at ($(origin)!0.42!(right)$);	
	\coordinate (Y1) at ($(origin)!0.9!(top)$); 	
	\coordinate (Y2) at ($(origin)!0.71!(top)$); 	
	\coordinate (Y3) at ($(origin)!0.42!(top)$); 	
	\coordinate (Y4) at ($(origin)!0.1!(top)$); 	
	\coordinate (P1) at (X4 |- Y3); 				
	\coordinate (P2) at (X3 |- Y4); 				
	\coordinate (P3) at (X1 |- Y4); 				
	\coordinate (P4) at (X5 |- Y2); 				
	\coordinate (P5) at (X2 |- Y4); 				
	\coordinate (P6) at (X1 |- Y1); 				
	\coordinate (P7) at (X6 |- Y4); 				
	\coordinate (A1) at ($(P4)!0.2!(P5)$);			
	\coordinate (A2) at ($(P1)!0.2!(P2)$);			
	\coordinate (A3) at ($(Y1)!0.9!(P7)$);			
	
	\draw (origin) -- ($(origin)-(0,3pt)$) node[anchor=north]{$0$};
	\draw (X1) -- ($(X1)-(0,3pt)$) node [anchor=north]{$1$};
	\draw (X2) -- ($(X2)-(0,3pt)$) node [anchor=north]{$1-r_F$};
	\draw (X3) -- ($(X3)-(0,3pt)$) node [anchor=north]{$1-r_F-r_D$};
	\draw (X4) -- ($(X4)-(0,3pt)$) node [anchor=north]{$\frac{1-r_F+r_D}{2+r_F+2r_D}$};
	\draw (X5) -- ($(X5)-(0,3pt)$) node [anchor=north]{$\frac{1-r_F}{2+r_F}$};
	\draw (X6) -- ($(X6)-(0,3pt)$) node [anchor=north, color=black!60]{$\frac{1-r_F}{2}$};
	\draw (Y1) -- ($(Y1)-(3pt,0)$) node [anchor=east]{$\frac{1}{r_F}$};
	\draw (Y2) -- ($(Y2)-(3pt,0)$) node [anchor=east]{$\frac{3}{2+r_F}$};
	\draw (Y3) -- ($(Y3)-(3pt,0)$) node [anchor=east]{$\frac{3}{2+r_F+2r_D}$};
	\draw (Y4) -- ($(Y4)-(3pt,0)$) node [anchor=east]{$1$};
	
	\draw[thick,blue, mark=*] plot (Y1) -- plot (P1) -- plot (P2) -- plot (P3); 
	\draw[thick,blue!30, mark=*] plot (Y1) -- plot (P7) -- plot (P3);
	\draw[thick, mark=square*, dashed, red] plot (Y1) -- plot (P4) -- plot(P5) -- plot (P3); 
	
	\draw[dotted] (Y2) -- (P4);
	\draw[dotted] (Y3) -- (P1);
	\draw[dotted] (Y4) -- (P2);
	\draw[dotted] (X5) -- (P4);
	\draw[dotted] (X4) -- (P1);
	\draw[dotted] (X3) -- (P2);
	\draw[dotted] (X2) -- (P5);
	\draw[dotted] (X1) -- (P3);
	\draw[dotted] (X6) -- (P7);
	
	\path[draw,dash dot,<->,red] (Y1) -- node[midway,above]{cloud + edge} (P6);
	\path[draw,<->,blue,dash dot] ($(Y4)-(0,5pt)$) -- node[midway,below]{cloud + edge + D2D} ($(P3)-(0,5pt)$);
	
	\draw[<-,red] ($(A1)+(45 : 5pt)$) -- ($(A1)+(45 : 2)$) node[anchor=west]{$r_D=0$};
	\draw[<-,blue] ($(A2)+(45 : 5pt)$) -- ($(A2)+(45 : 2)$) node[anchor=west]{$0<r_D<\frac{1-r_F}{2}$};
	\draw[<-,blue!60] ($(A3)+(45 : 5pt)$) -- ($(A3)+(45 : 2.2)$) node[anchor=west]{$r_D\geq\frac{1-r_F}{2}$};

	
\end{tikzpicture}